\documentclass[aps,PhysRevB,reprint,superscriptaddress,showpacs]{revtex4-2}

\usepackage[T1]{fontenc}
\usepackage{mathptmx}
\usepackage{graphicx}
\usepackage[version=4]{mhchem}
\usepackage{textcomp}
\usepackage{dcolumn}
\usepackage{bm}
\usepackage{esdiff}
\usepackage{natbib}
\usepackage[colorlinks=true,citecolor=blue,linkcolor=blue,urlcolor=blue,pdfhighlight =/O]{hyperref}
\begin{document}
\title{Magnetic field-tuned quantum criticality in optimally electron-doped cuprate thin films}

\author{Xu Zhang}
\altaffiliation{These authors contributed equally to this work.}
\affiliation{Beijing National Laboratory for Condensed Matter Physics, Institute of Physics, Chinese Academy of Sciences, Beijing 100190, China}
\affiliation{School of Physical Sciences, University of Chinese Academy of Sciences, Beijing 100049, China}

\author{Heshan Yu}
\altaffiliation{These authors contributed equally to this work.}
\affiliation{Beijing National Laboratory for Condensed Matter Physics, Institute of Physics, Chinese Academy of Sciences, Beijing 100190, China}
\affiliation{School of Physical Sciences, University of Chinese Academy of Sciences, Beijing 100049, China}
\affiliation{Department of Materials Science and Engineering, University of Maryland, College Park, Maryland 20742, US}

\author{Qihong Chen}
\affiliation{Beijing National Laboratory for Condensed Matter Physics, Institute of Physics, Chinese Academy of Sciences, Beijing 100190, China}
\affiliation{School of Physical Sciences, University of Chinese Academy of Sciences, Beijing 100049, China}

\author{Runqiu Yang}
\affiliation{Quantum Universe Center, Korea Institute for Advanced Study, Seoul 130-722, Korea}
\affiliation{Center for Joint Quantum Studies and Department of Physics, School of Science, Tianjin University, Yaguan Road 135, Jinnan District, 300350 Tianjin, China}

\author{Ge He}
\affiliation{Beijing National Laboratory for Condensed Matter Physics, Institute of Physics, Chinese Academy of Sciences, Beijing 100190, China}
\affiliation{School of Physical Sciences, University of Chinese Academy of Sciences, Beijing 100049, China}

\author{Ziquan Lin}
\affiliation{Wuhan National High Magnetic Field Center (WHMFC), Huazhong University of Science and Technology, Wuhan 430074, China}

\author{Qian Li}
\affiliation{Beijing National Laboratory for Condensed Matter Physics, Institute of Physics, Chinese Academy of Sciences, Beijing 100190, China}

\author{Jie Yuan}
\affiliation{Beijing National Laboratory for Condensed Matter Physics, Institute of Physics, Chinese Academy of Sciences, Beijing 100190, China}
\affiliation{School of Physical Sciences, University of Chinese Academy of Sciences, Beijing 100049, China}
\affiliation{Songshan Lake Materials Laboratory, Dongguan, Guangdong 523808, China}

\author{Beiyi Zhu}
\affiliation{Beijing National Laboratory for Condensed Matter Physics, Institute of Physics, Chinese Academy of Sciences, Beijing 100190, China}
\affiliation{School of Physical Sciences, University of Chinese Academy of Sciences, Beijing 100049, China}

\author{Liang Li}
\affiliation{Wuhan National High Magnetic Field Center (WHMFC), Huazhong University of Science and Technology, Wuhan 430074, China}

\author{Yi-feng Yang}
\author{Tao Xiang}
\affiliation{Beijing National Laboratory for Condensed Matter Physics, Institute of Physics, Chinese Academy of Sciences, Beijing 100190, China}
\affiliation{School of Physical Sciences, University of Chinese Academy of Sciences, Beijing 100049, China}

\author{Rong-Gen Cai}
\affiliation{CAS Key Laboratory of Theoretical Physics, Institute of Theoretical Physics, Chinese Academy of Sciences, Beijing 100190, China}
\affiliation{School of Physical Sciences, University of Chinese Academy of Sciences, Beijing 100049, China}

\author{Anna Kusmartseva}
\affiliation{Department of Physics, Loughborough University, Loughborough LE11 3TU, United Kingdom}

\author{F. V. Kusmartsev}
\affiliation{Department of Physics, Loughborough University, Loughborough LE11 3TU, United Kingdom}
\affiliation{College of Art and Science, Khalifa University, PO Box 127788, Abu Dhabi, UAE}

\author{Jun-Feng Wang}
\affiliation{Wuhan National High Magnetic Field Center (WHMFC), Huazhong University of Science and Technology, Wuhan 430074, China}

\author{Kui Jin}\email[]{kuijin@iphy.ac.cn}
\affiliation{Beijing National Laboratory for Condensed Matter Physics, Institute of Physics, Chinese Academy of Sciences, Beijing 100190, China}
\affiliation{School of Physical Sciences, University of Chinese Academy of Sciences, Beijing 100049, China}
\affiliation{Songshan Lake Materials Laboratory, Dongguan, Guangdong 523808, China}

\date{\today}

\begin{abstract}
Antiferromagnetic (AF) spin fluctuations are commonly believed to play a key role in electron pairing of cuprate superconductors. In electron-doped cuprates, it is still in paradox about the interplay among different electronic states in quantum perturbations, especially between superconducting and magnetic states. Here, we report a systematic transport study on cation-optimized La$_{2-x}$Ce$_x$CuO$_{4\pm\delta}$ ($x=0.10$) thin films in high magnetic fields. We find an AF quantum phase transition near 60 T, where the Hall number jumps from $n_{\mathrm{H}}=-x$ to $n_{\mathrm{H}}=1-x$, resembling the change of $n_{\mathrm{H}}$ at the AF boundary ($x_{\mathrm{AF}}=0.14$) tuned by Ce doping. In the AF region a spin dependent state manifesting anomalous positive magnetoresistance is observed, which is closely related to superconductivity. Once the AF state is suppressed by magnetic field, a polarized ferromagnetic state is predicted, reminiscent of the recently reported ferromagnetic state at the quantum endpoint of the superconducting dome by Ce doping. The magnetic field that drives phase transitions in a similar but distinct manner to doping thereby provides a unique perspective to understand the quantum criticality of electron-doped cuprates. 
\end{abstract}

\maketitle

\section{Introduction}

High-temperature superconductivity commonly lies within a special region as a function of tuning parameters (e.g., chemical substitutions), where quantum criticality governs the essential physics \cite{RevModPhys.77.721,RevModPhys.82.2421,RevModPhys.84.1383,RevModPhys.87.855,keimer2015quantum}. In the electron-doped cuprates, antiferromagnetic (AF) order emerging from the Mott insulating phase competes with superconductivity. Advanced techniques like muon spin rotation and neutron scattering experiments unveiled that the long-range AF order vanishes before the appearance of superconductivity \cite{saadaoui2015phase,motoyama2007spin}. However, the reconstruction of Fermi surface (FS) underneath the superconducting dome observed by electrical transport \cite{jin2011link,Mandal5991} and angle resolved photoemission spectroscopy (ARPES) \cite{He3449,RN159} measurements, requires an extended AF state or spin density wave into the overdoped region. Although many works give evidence to support the AF scenario \cite{PhysRevLett.94.057005,PhysRevB.72.214506,Mandal5991}, the origin of the reconstruction remains controversial. Very recently, ferromagnetic (FM) \cite{Sarkar532} order was reported in the electron-doped cuprates, which adds more complexity to this system. Superconductivity is accompanied by the evolution of magnetic ordered states from birth to death. A deep understanding of the interplay between superconducting and magnetic states is needed.

\begin{figure}[htbp]
\includegraphics[width=\columnwidth]{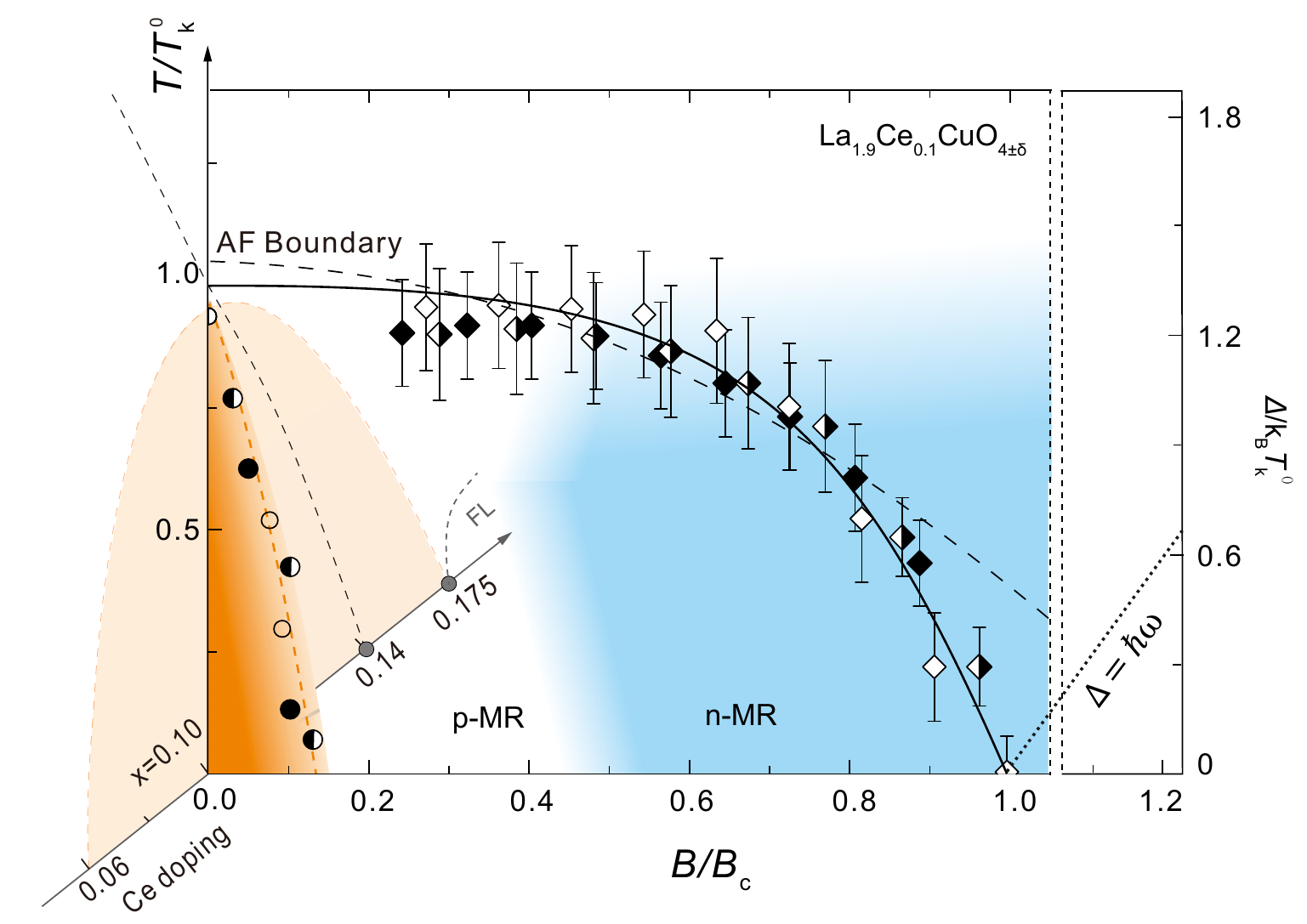}
\caption{\label{Fig1}Phase diagram of LCCO obtained from magneto-electrical transport measurements. The side panel illustrates the Ce doping dependence of the superconducting dome. With increasing doping, LCCO exhibits a maximum $T_{\mathrm{c}}$ at $x=0.10$. The main panel shows the field-temperature ($B$-$T$) diagram of $x=0.10$ samples (S1, S2, and S3). The coordinate axes are normalized by two characteristic quantities $B_{\mathrm{c}}$ and $T_{\mathrm{k}}^0$, respectively. Here, $B_{\mathrm{c}}$ is obtained by extrapolating the boundary of canted AF state to zero-temperature limit (for the three samples, $B_{\mathrm{c}}\sim$ 62, 55, and 52 T). $T_{\mathrm{k}}^0$ corresponds to the temperature of Fermi surface reconstruction in zero-field limit (for the three samples, $T_{\mathrm{k}}^0\sim$ 32, 27, and 26 K), manifesting as a `kink' in the Hall coefficient (see Fig. \ref{Fig2}). The normalized boundaries of three samples can be fitted by the holographic model (black solid line), which also predicts an energy gap (dotted line) above $B_{\mathrm{c}}$. Three regions can be identified in the $B$-$T$ diagram, i.e., $n$-MR (blue), $p$-MR (white) and superconducting (orange) regions. The dashed line is mean-field fitting, which deviates data near the quantum critical point $B_{\mathrm{c}}$.}
\end{figure}

Magnetic field as a parameter of quantum phase transitions can tune both the superconductivity and magnetic ordered states, by which the evolution of spin-dependent states can been obtained. Among all electron-doped cuprates, La$_{2-x}$Ce$_x$CuO$_{4\pm\delta}$ (LCCO) is a unique one that exhibits a complete superconducting dome with Ce doping, as illustrated in Fig. \ref{Fig1}. Upon Ce substitution of La, the superconductivity emerges at $x=0.06$ and reaches the maximum transition temperature ($T_{\mathrm{c}}$) at $x=0.10$, where it intersects with the boundary of the AF state (or spin density wave) \cite{Mandal5991,ZHANG201618,PhysRevB.96.155449}. At this point, the magnetic field can act as a perturbation in energy to tip the balance between the AF and superconducting states. Therefore, LCCO provides a good platform to study the electronic states as a function of magnetic field. 

Here, we performed systematic magnetotransport measurements up to 55 T on optimally doped LCCO thin films ($x=0.10$) and established an intriguing field-tuned phase diagram. As summarized in Fig. \ref{Fig1}, the field-temperature ($B$-$T$) panel reveals three phase transitions at zero-temperature limit: (i) a superconducting phase transition at $\sim$ 9 T; (ii) an AF quantum phase transition at a characteristic field $B_{\mathrm{c}}$; and (iii) a plausible topological phase transition that coexists with the short-range AF order in between, signified by the crossover of resistivity from an anomalous enhancement (positive magnetoresistance, $p$-MR) to an reduction (negative magnetoresistance, $n$-MR) with increasing magnetic field. The $p$-MR region, which isolates the superconducting phase from the $n$-MR region can extend to $\frac{1}{2}B_{\mathrm{c}}$. In addition, samples with different $T_{\mathrm{c}}$'s (by slightly tuning oxygen) show a universal AF-state boundary that can be described by the holographic model (the black dashed line). This model also predicts a soft spin gap above $B_{\mathrm{c}}$ (dotted line at the bottom right corner). Thereafter, we will demonstrate the AF phase transition based on our magnetotransport results and explore the origin of the $p$-MR. 

\section{Experiments}

The $c$-axis-oriented LCCO ($x=0.10$) films were deposited on the $(00l)$-oriented SrTiO$_3$ substrates by a pulsed laser deposition system. The thicknesses of the films were measured by scanning electron microscope: $d=230$ nm (S1, S2) and 110 nm (S3, S4). Detailed transport measurements were carried out on four LCCO film samples with different $T_{\mathrm{c}0}$'s: 25.6 K (S1), 25.3 K (S2), 24.1 K (S3) and 23.1 K (S4) by slightly tuning the oxygen content during the film deposition. Here, $T_{\mathrm{c}0}$ is the critical temperature at which the resistance reaches zero. Magnetotransport measurements in field up to 58 T were performed using a non-destructive pulsed magnet with a pulsed duration of 60 msec at Wuhan National High Magnetic Field Center. Magnetoresistance (MR) and Hall resistance were measured simultaneously with a typical Hall-bar configuration. Data for the up- and down-sweeping of the pulse field were in good agreement, thus we could exclude the heating effect of the sample by the eddy current. Measurements with both positive and negative field polarities were performed for all samples and measuring temperatures to eliminate the effect of contact asymmetries.

\section{Results}

\begin{figure*}[ht!]
	\centering
	\includegraphics[width=\linewidth]{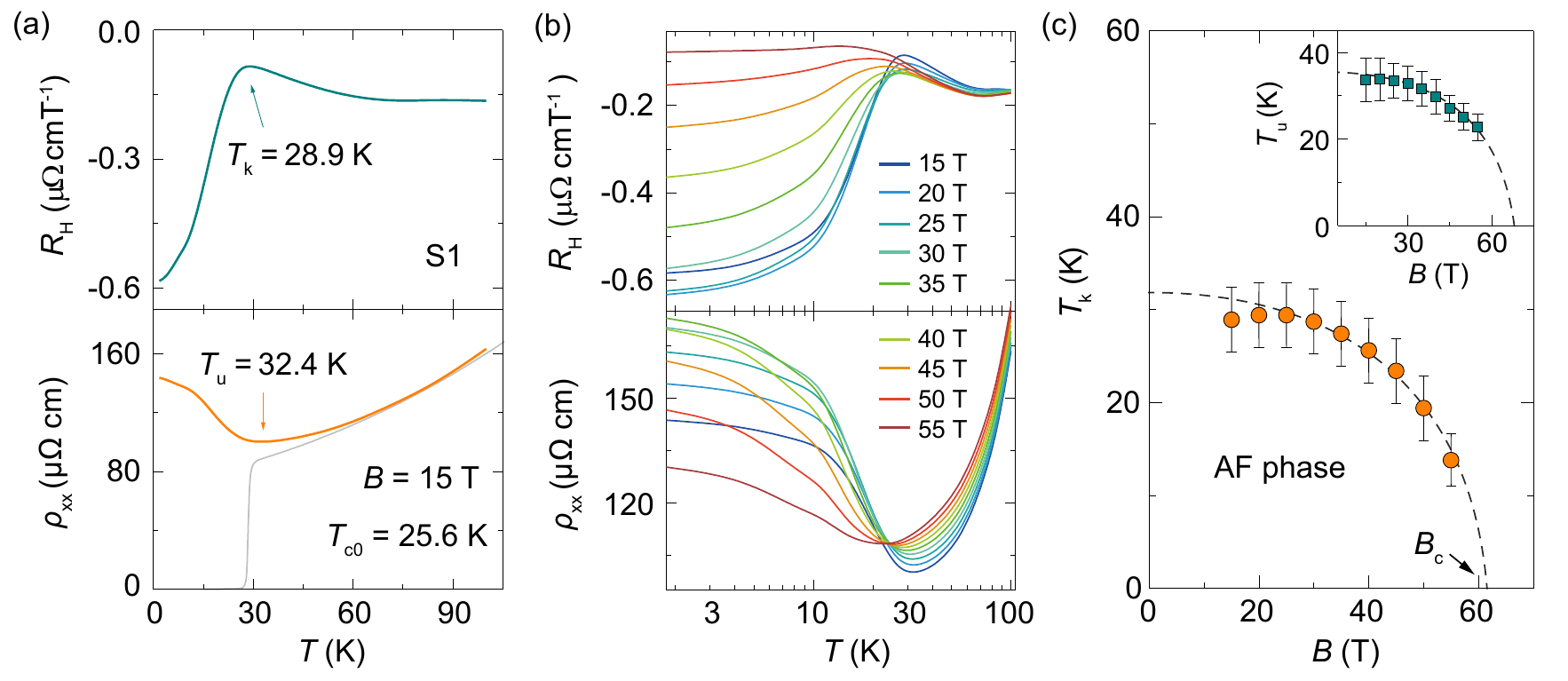}
	\caption{\label{Fig2}Hall `kink' and resistivity `upturn' in LCCO. (a) Temperature dependence of the $R_{\mathrm{H}}$ and $\rho_{xx}$ at 15 T for sample S1. $T_{\mathrm{k}}$ ($T_{\mathrm{u}}$) is extracted from the maximum (minimum) of $R_{\mathrm{H}}$ ($\rho_{xx}$) curves. (b) Temperature dependence of the Hall coefficient (top panel) and resistivity (bottom panel) at different fields up to 55 T for sample S1. (c) The magnetic field dependence of $T_{\mathrm{k}}$. Inset: The magnetic field dependence of $T_{\mathrm{u}}$.}
\end{figure*}

Relatively low upper critical field of electron-doped cuprates (e.g., typically $<10$ T for LCCO) facilitates the study of phase transitions via electrical transport. Figure \ref{Fig2}(a) shows the temperature dependence of the Hall coefficient $R_{\mathrm{H}}$ ($R_{\mathrm{H}}(T,B)=\rho_{xy}(T)/B$) and the resistivity $\rho_{xx}(T)$ at $B=15$ T. As the superconducting state is destroyed by the field, we observe a `kink' in the $R_{\mathrm{H}}$ curve, namely, the Hall coefficient changes gently at high temperature but drops quickly below a characteristic temperature $T_{\mathrm{k}}$. Here, $T_{\mathrm{k}}$ is the temperature at which the Hall coefficient reaches the maximum ($T_{\mathrm{k}}\sim 29\pm 3$ K for sample S1). Concurrently, an upturn in the resistivity appears at the temperature $T_{\mathrm{u}}\sim 32\pm4$ K, that is, from $d\rho/dT>0$ to $d\rho/dT<0$ as the temperature decreases. In electron-doped cuprates, the `kink' in $R_{\mathrm{H}}(T)$ and the corresponding `upturn' in $\rho_{xx}(T)$ are commonly regarded as signatures of FS reconstruction \cite{PhysRevB.96.155449,PhysRevB.80.012501}.

Figure \ref{Fig2}(b) shows the $R_{\mathrm{H}}(T)$ and $\rho_{xx}(T)$ curves of S1 in different magnetic fields. Remarkably, both the kink in $R_{\mathrm{H}}(T)$ and the upturn in $\rho_{xx}(T)$ fade away at high magnetic field, and $T_{\mathrm{k}}$ ($T_{\mathrm{u}}$) gradually shifts toward lower temperatures [Fig. \ref{Fig2}(c)]. In electron-doped cuprates, the FS reconstruction is usually attributed to the AF ordering \cite{PhysRevB.96.155449,PhysRevB.80.012501}. The evolution of Hall coefficient with Ce doping is consistent with theory, which considers a two-dimensional system undergoing a spin density wave instability \cite{PhysRevB.72.214506}. And near the quantum critical point, the AF fluctuation is detected \cite{Mandal5991}. In our results, the `kink' and `upturn' behaviors can be tuned by magnetic field, which provides strong evidence that the Fermi surface reconstruction origin from spin instead of charge. Since the Hall signal is more sensitive to carrier number, we will use the $T_{\mathrm{k}}(B)$ curve to track the energy scales of temperature and field for the AF transition. By extrapolating the $T_{\mathrm{k}}(B)$ to zero temperature, one can obtain a critical field $B_{\mathrm{c}}$ where the kink is suppressed by the field in quantum limit (i.e., $T=0$ K). Also, that $T_{\mathrm{k}}^0=T_{\mathrm{k}}(B=0)$ reveals the energy scale of classic AF phase transition at zero-field limit. Using normalized temperature and field, i.e., $T/T_{\mathrm{k}}^0$ and $B/B_{\mathrm{c}}$, we are able to scale these two quantities in shaping the electronic phase diagram of LCCO as shown in Fig. \ref{Fig1}. It is interesting that the boundaries of the AF state for samples S1-S3 almost overlap with each other, suggesting a universal scaling between temperature and magnetic field.

\begin{figure}[htbp]
\includegraphics[width=\columnwidth]{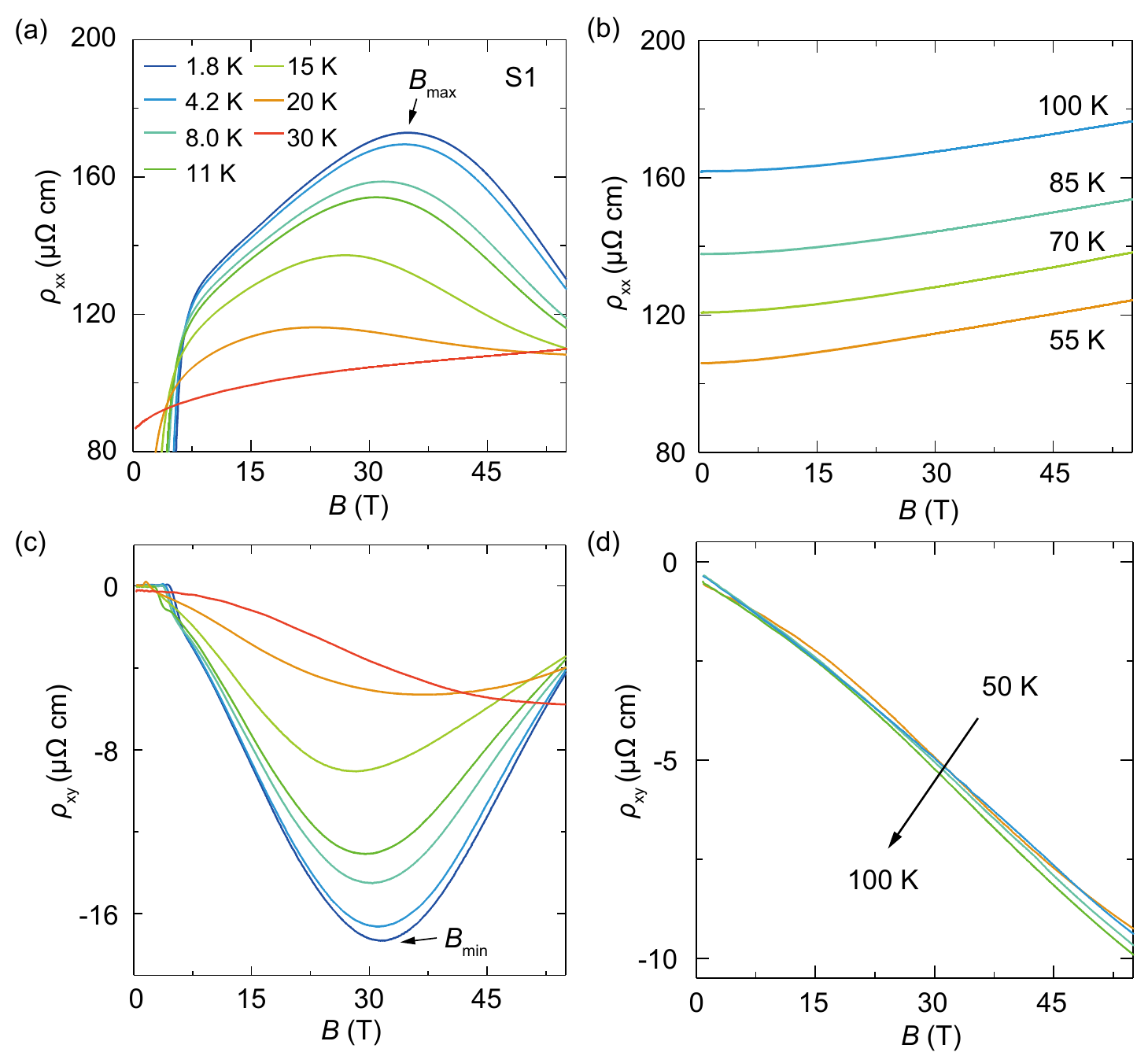}
\caption{\label{Fig3}Longitudinal resistivity and Hall resistivity of sample S1 from 1.8 to 100 K. (a, b) The field-dependent longitudinal resistivity displays non-monotonous behavior below 30 K. The $B_{\max}$ marks the characteristic magnetic fields where the resistivity reaches the maximum. Above 30 K, magnetoresistance shows linearity at high magnetic fields. (c, d) With decreasing temperature, there is a switching from linear to non-linear behavior. The minimum in $\rho_{xy}$ curves corresponding to $B_{\max}$ also disappears above 30 K.}
\end{figure}

In Fig. \ref{Fig2}(b), different $\rho_{xx}(T)$ curves overlap at low temperatures, indicating that the resistivity is non-monotonically dependent on the magnetic field. Such non-monotonic behavior can be better viewed in isothermal $\rho_{xx}(B)$ curves. As shown in Fig. \ref{Fig3}(a), a crossover from $p$-MR ($d\rho/dB>0$) to $n$-MR ($d\rho/dB<0$) occurs below 30 K, which becomes more prominent at lower temperatures. We define a characteristic magnetic field $B_{\max}$ that gives the maximum resistivity, which moves to higher field with decreasing temperature. Simultaneously, the isothermal Hall resistivity curves $\rho_{xy}(B)$ display minima at low temperatures as shown in Fig. \ref{Fig3}(c). Above 30 K, the crossover disappears and the resistance is always enhanced with magnetic field ($p$-MR) [Fig. \ref{Fig3}(b)]. The simultaneous appearance of these two extremum points indicates that there is a competition between two electronic states, which is strongly dependent with temperature.

\section{Discussions}

\subsection{A jump of Hall number in AF phase transition}

\begin{figure}[htbp]
\includegraphics[width=\columnwidth]{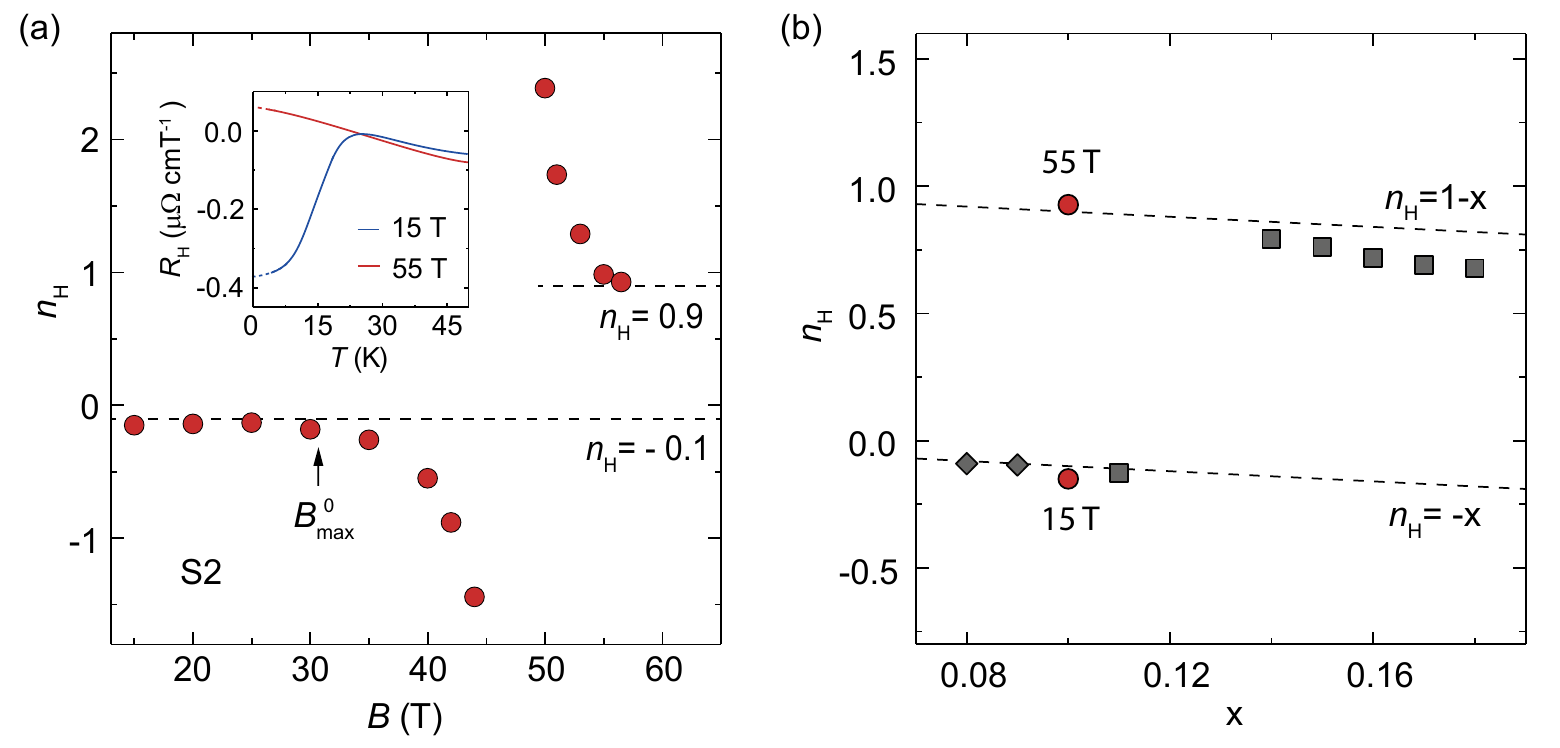}
\caption{\label{Fig4}The Hall number and AF transition. (a) The Hall number $n_{\mathrm{H}}$ as a function of magnetic field. The $n_{\mathrm{H}}$ deviates from $-0.1$ at $B_{\max}^0$. Inset: Temperature dependence of the Hall coefficient at 15 and 55 T for sample S2. (b) The Hall number $n_{\mathrm{H}}$ as a function of Ce doping (square from Ref. \cite{PhysRevB.96.155449} and diamond from Ref. \cite{PhysRevB.80.012501}). The upper dashed line marks $n_{\mathrm{H}}=1-x$; the lower dashed line marks $n_{\mathrm{H}}=-x$.}
\end{figure}

To describe the universal quantum phase transition more quantitatively, we extract the Hall number, $n_{\mathrm{H}}=\frac{V}{eR_{\mathrm{H}}(0)}$, which reflects the average carrier doping per Cu atom. Here, $e$ is the electron charge; $V$ is the volume per Cu atom in the CuO$_2$ planes and $R_{\mathrm{H}}(0)$ represents the Hall coefficient extrapolated to zero temperature. Figure \ref{Fig4}(a) shows the field dependent $n_{\mathrm{H}}$ of sample S2 (the $B_{\mathrm{c}}$ of this sample is slightly lower than that of S1, and thereby can be accessed in the experiments). At low fields, the Hall number $n_{\mathrm{H}}$ equals $-0.1$; at high fields, it jumps to a positive value and tends to saturate at 55 T, where $n_{\mathrm{H}}=0.9$. This jump brings to mind the Ce doping dependence of Hall number $n_{\mathrm{H}}(x)$, which follows a similar change from $-x$ to $1-x$ at the quantum critical point (QCP) ($x_{\mathrm{AF}}=0.14$, as seen in Fig. \ref{Fig1}) \cite{PhysRevB.96.155449}. As shown in Fig. \ref{Fig4}(b), the values of $n_{\mathrm{H}}(B)$ at 15 and 55 T match well with the data extracted from Refs. \cite{PhysRevB.96.155449,PhysRevB.80.012501}, suggesting a field-induced FS reconstruction. ARPES experiments have revealed a large hole-like pocket around $(\pi,\pi)$ in overdoped Nd$_{2-x}$Ce$_x$CuO$_4$ (NCCO), which can be reconstructed to small electron pockets once AF state enters at lower Ce doping levels \cite{PhysRevB.75.224514,PhysRevLett.88.257001}. The resemblance between $n_{\mathrm{H}}(B)$ and $n_{\mathrm{H}}(x)$ suggests the recovery of a large hole pocket FS when the AF order is suppressed.

\subsection{The fitting of AF boundary}

In order to understand the AF boundary in a quantitative way, two methods including the mean-field theory and holographic model are considered. The analysis of a magnetic system in the frame work of the mean-field approach involves its thermodynamic potential given by Lev Landau \cite{Landau}. In AF phase, the mean field order parameter $L$ for antiferromagnetic order is equal to $\sqrt{a(T-T_{\mathrm{c}})/2b+\beta/2b}$ and the critical temperature in this case is determined by the equation $a(T-T_{\mathrm{N}}^0)\pm\beta/2=0$. At weak field limit, the Neel temperature is given by $T_{\mathrm{N}}(B)=T_{\mathrm{N}}^0-D'B^2/a-\beta/2a$. Here, $a$, $b$ and $D'$ are parameters which can be determined from a comparison with experiments. The $T_{\mathrm{N}}^0$ is the Neel temperature at zero-field limits. The fitting curve based on the mean-field theory is shown in Fig. \ref{Fig1} (dashed line). At low field region, the Neel temperature of the AF critical transition depends quadratically on the magnetic field, while deviates fitting curve near critical magnetic field. Therefore, the mean-field theory cannot describe the phase transition, nor predict the quantum critical point accurately in our system.

For the strongly correlated system, a new method, called holographic model is developed from the string theory and has been widely applied to study various strongly correlated phenomena, especially the AF phase transition \cite{PhysRevD.92.046005,PhysRevD.92.086001}. We attempt to apply this model to our system. The boundary of AF state for three samples can be well fitted by the holographic model in the whole magnetic field range (black solid line in Fig. \ref{Fig1}, see details in Appendix). In the limit of the small magnetic field ($B\ll B_{\mathrm{c}}$), the holographic model also gives a quadratic relation with magnetic field same as the mean field calculation. However, when the magnetic field is large, a completely different violation of the hyper-scaling equation is obtained. The function of $T_{\mathrm{N}}$ has an exact asymptotic behavior as $T_{\mathrm{N}}/T_{\mathrm{N}}^0\sim(1-B/B_{\mathrm{c}})$. In this case the Neel temperature, denoted by $T_{\mathrm{k}}$ in this paper, vanishes in magnetic field in a highly nontrivial way described by this equation.

Successfully utilizing this model not only provides another piece of evidence for the AF quantum phase transition, but also predicts a polarized FM state above $B_{\mathrm{c}}$, with an excited energy gap $\Delta$ satisfying: $$\Delta\sim k_{\mathrm{B}}T_{\mathrm{k}}^0\left(\frac{B}{B_{\mathrm{c}}}-1\right),\quad 0<\frac{B}{B_{\mathrm{c}}}-1\ll 1.$$Perhaps not coincidently, a recent work reported the existence of a FM order right beyond the endpoint of the superconducting dome in LCCO \cite{Sarkar532}. The field-tuned phase diagram of optimally doped LCCO discloses one AF-FM QCP at $B_{\mathrm{c}}$. In contrast, the Ce doping phase diagram shows two QCPs, i.e., $x_{\mathrm{AF}}=0.14$ and $x_{\mathrm{FM}}=0.175$. Whether the involvement of superconductivity and its interaction with the AF state break the AF-FM QCP into two is an interesting issue worthy of future study. 

\subsection{Abnormal $p$-MR}

\begin{figure}[htbp]
\includegraphics[width=\columnwidth]{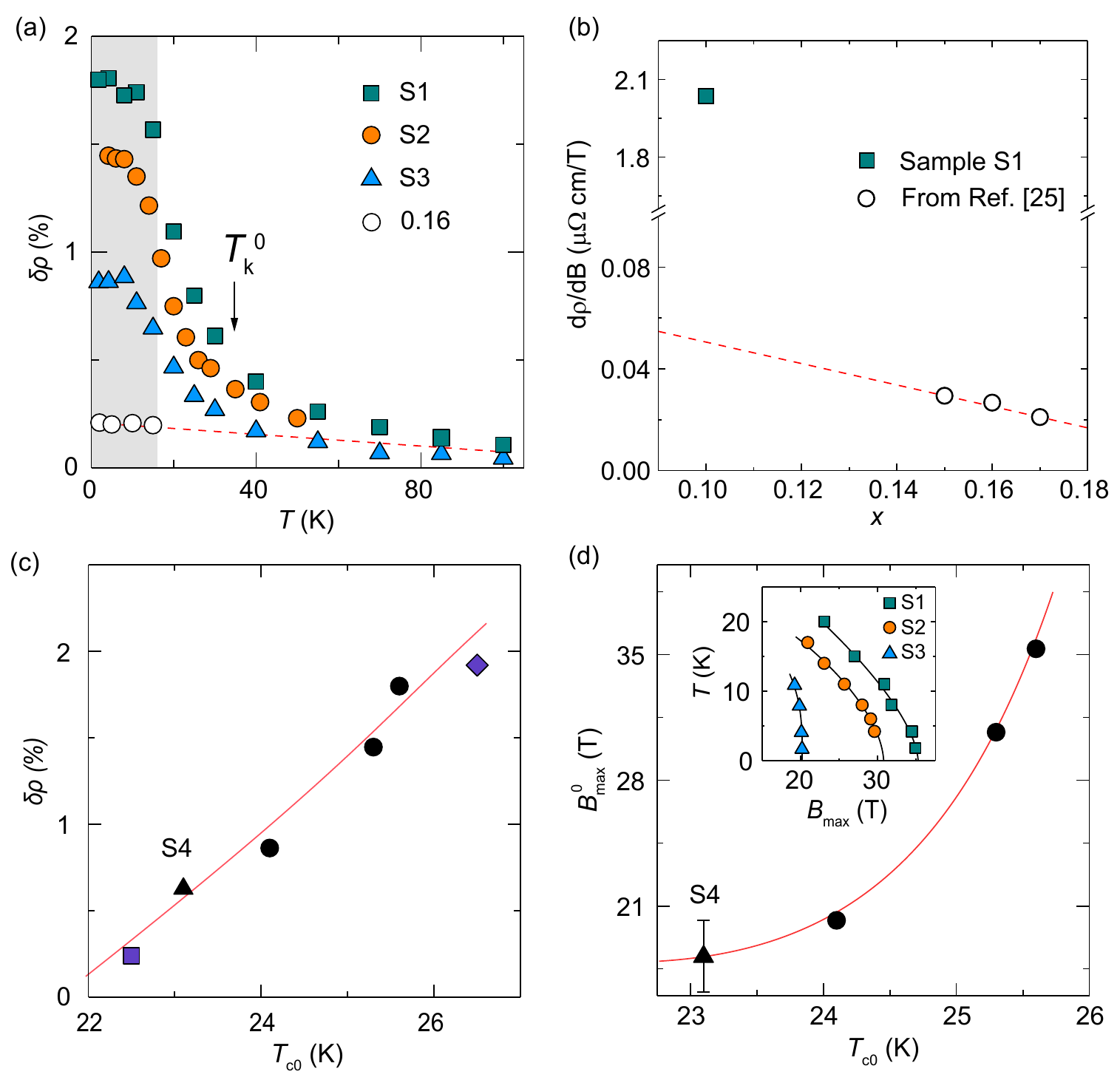}
\caption{\label{Fig5}The characteristic quantities of $p$-MR related to $T_{\mathrm{c}}$ (a) The  $\delta\rho=\left.\frac{d\rho(B)}{dB}\frac{1}{\rho(B)}\right|_{B=12\,\mathrm{T}}$ as a function of temperature for different LCCO samples (white dots for $x=0.16$ from Ref. \cite{Sarkareaav6753}). (b) The slop of $\rho_{xx}(B)$ for LCCO with different doping (square at 1.8 K, circles at 400 mK from Ref. \cite{Sarkareaav6753}). (c) The correlation between $\delta\rho$ and $T_{\mathrm{c}0}$ for different LCCO samples (filled circles at 1.8 K, triangle at 8 K, square at 2 K from Ref. \cite{PhysRevB.77.172503}, diamond at 4.5 K from Ref. \cite{PhysRevB.78.174521}). (d) The correlation between $B_{\max}^0$ and $T_{\mathrm{c}0}$. Inset: The temperature dependence of $B_{\max}$ for sample S1. $B_{\max}^0$ is obtained from the epitaxial curve to zero temperature.}
\end{figure}

The $n$-MR is common in AF state \cite{PhysRevLett.94.057005,PhysRevB.77.172503,Yu2017}. In our study, the starting field of $n$-MR ($B_{\max}$) is the same as the characteristic field where $n_{\mathrm{H}}$ deviates from $-0.1$ as shown in Fig. \ref{Fig4}(a). This indicates the $n$-MR is closely tied to the modulation of the AF spin polarization by the magnetic field. Therefore, the $n$-MR can be naturally explained: Carrier hopping from Cu sites to its nearest neighbors is prohibited in some extend by the AF spin configuration, but it is possible that the magnetic field deflects spins of the short-range AF order and provides a polarized FM channel (mimicking the canted AF state), which subsequently enhances the conductivity and causes the $n$-MR effect. The $p$-MR at low temperatures seems unusually large compared to the normal MR. We take a definition of $\delta\rho(B)=\frac{d\rho(B)}{dB}\frac{1}{\rho(B)}$ to evaluate the magnitude of the anomalous $p$-MR. Figure \ref{Fig5}(a) shows $\delta\rho(12\,\mathrm{T})$ as a function of temperature for samples S1-S3. It is almost temperature independent below 15 K, and decreases substantially as the temperature rises. Above the onset temperature of the AF state, it drops more than 90 percent and becomes comparable to the normal MR of $x=0.16$ sample without an AF order \cite{Sarkareaav6753} (white dots). 

For optimally-doped LCCO, the upper critical field is about $\sim 9$ T. Previous experiments have clearly shown that superconducting fluctuations in electron-doped cuprates are weak above the critical temperature/field compared with their hole-doped counterparts \cite{PhysRevB.73.024510}. For instance, the large Nernst voltage due to superconducting fluctuations will be quickly wiped out once the field is a few Tesla higher than the upper critical field \cite{PhysRevB.76.174512,PhysRevB.97.014522}. Therefore, superconducting fluctuations are not expected to contribute to the $p$-MR up to 35 T.

Recently, charge order has been disclosed in electron-doped cuprates \cite{daSilvaNeto282,daSilvaNetoe1600782,RN160}. Matsuoka {\it et al.} observed an enhancement of $T_{\mathrm{u}}$ (upturn in resistivity) and attribute it to charge order correlations \cite{PhysRevB.98.144506}. However, on one hand, the strength of the charge order is even stronger in the overdoped region \cite{daSilvaNetoe1600782,RN160}, but the anomalous $p$-MR is not observed in overdoped LCCO. On the other hand, the $\delta\rho$ for LCCO in this work as well as the data from previous studies \cite{PhysRevB.77.172503,PhysRevB.78.174521} shows a positive correlation with $T_{\mathrm{c}0}$ in a range from 22 to 26 K [Fig. \ref{Fig5}(c)]. The positive correlation also does not support charge order as the origin of the unusual $p$-MR effect, since it competes with the superconductivity \cite{chang2012direct,Ghiringhelli821}. 

Note that the greatly enhanced $\delta\rho$ at low temperatures occurs in the AF region, suggesting an unambiguous link between the $p$-MR effect and the AF order. It has been reported that the FS reconstruction induced by the AF order leads to the coexistence of electron and hole pockets, which may also result in a $p$-MR behavior \cite{PhysRevB.76.174512,li2019hole}. However, in the $p$-MR region of our study, the Hall number ($n_{\mathrm{H}}=-0.1$) is consistent with the nominal value of electron carriers, suggesting a dominant contribution from the electron pocket. This result agrees with a previous study \cite{PhysRevB.69.024504}, which demonstrates a similar effect between temperature and doping in tuning the FS. By decreasing the temperature/doping, electron pocket is supposed to dominate the transport eventually at low $T$ or $x$, challenging the explanation of two-band-driven $p$-MR.

\subsection{Topological order}

\begin{figure}[htbp]
\includegraphics[width=\columnwidth]{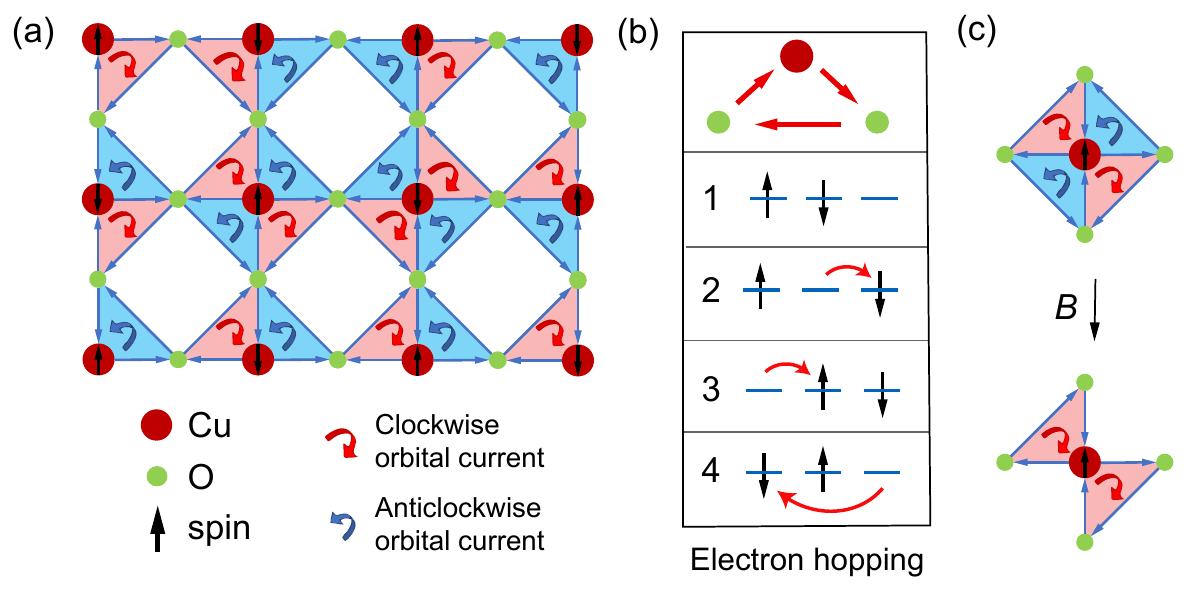}
\caption{\label{Fig6}The schematic description of the topological order. (a) The topological order is best visualized by considering a Cu-O-O triangular plaquette on the Cu-O plane. On the plaquette the Cu$^{2+}$-ion has a pre-localized spin and two neighboring oxygens. Clockwise and anticlockwise $\pi$-orbital currents or moments are shown in pink and blue, respectively. (b) The sequence tunneling of a doped electron. Such state is unstable and this instability can lead to the formation of spontaneous local orbital current. (c) A schematic depiction of the dissociation of the vortex-antivortex pairs (or clockwise and anticlockwise orbital currents) by magnetic field. The magnetic field is applied perpendicular to the plane of the quadruple moments. The field polarizes orbital currents of a particular chirality and thereby causes an unbinding of the vortex-antivortex pairs when the limit is approached. This leads to a Beresinki-Kosterlitz-Thouless like phase transition.}
\end{figure}

A topological order coexisting with short-range AF order \cite{PhysRevX.8.021048,ScheurerE3665,Sachdev_2018} has been proposed recently to understand the energy gap in overdoped NCCO in the ARPES \cite{He3449} and Shubnikov-de Haas oscillation \cite{PhysRevLett.105.247002} studies. Here, we consider this scenario and propose a reasonable configuration of topological order which consistently explains the $p$-MR observed in our experiment. The presence of this topological order, based on the displacement of doped electron among the Cu$^{2+}$ and its neighboring oxygen sites, prevents the spins with $\pi$-fluxes from being realigned by the magnetic field [as shown in Fig. \ref{Fig6}(a)]. Each triangular plaquette may have a positive or negative chirality, which is associated with the orientation of the spin-orbital currents arising around the doped Cu site. Electrons doped into copper Cu$^{2+}$-site tend to move to the neighboring oxygens due to strong on-site Hubbard $U$. A possible sequence of electron hopping along the shortest triangular trajectory is depicted on right panel of Fig. \ref{Fig6}(b) (lines 1-4). As a result of these tunneling sequences, we have the same charge but different spin configurations on the triangular O-Cu-O, which constitutes a $\pi$-phase change. Only by moving the electron twice around the same triangular path is the total phase $2\pi$ and the original state recovered. Therefore, this specific topological order supports orbital currents or moments, related to gauge $Z_2$ field. 

In each quadruplet there are four ordered plaquettes: two have positive chirality and two have negative chirality. There is a local $C_2$ symmetry similar to a $d$-wave-like state. Here, the sign is simply an indication of the spin-orbital current associated with each plaquettes. Each quadruplet forms four $\pi$-fluxons with zero topological charge and is nearly homogeneously distributed around CuO$_2$ plane. The $\pi$-flux is created together with spin and orbital currents. That spin-orbit coupling keeps the in-plane orientation of the spin involved in a particular $\pi$-flux. This naturally impedes the deflection of the spins and leads to a $p$-MR behavior. 

The magnetic field may induce an imbalance between the plaquettes of the positive and negative chirality. As the magnetic field increases, the plaquettes of one particular chirality are enhanced, while the plaquettes of the opposing chirality are suppressed. This leads to a decrease of the binding energy of $\pi$-fluxes in the quadruplet configuration. As a result, at some critical magnetic field the topological $\pi$-fluxons are decoupled from quadruplets and topological phase transition occurs as shown in Fig. \ref{Fig6}(c). The de-coupling of the $\pi$-fluxons is similar to the unbinding of vortices and anti-vortices in the Beresinski-Kosterlitz-Thouless (BKT) transition. Note that in the proposed topological order only the spins involved in the formation of $\pi$-fluxons quadruplets are not canted and contribute to $p$-MR properties. When all $\pi$-fluxons are unbound, the quadruplets vanish and the screening orbital currents disappear. Consequently, we see a gradual transition from $p$-MR to $n$-MR as the magnetic field increases.

\subsection{The relation between $p$-MR and superconductivity}

As aforementioned, the $\delta\rho$ is generally larger for samples with higher $T_{\mathrm{c}}$ in a window of 22 to 26 K by slightly tuning the oxygen of optimally Ce-doped samples [Fig. \ref{Fig5}(c)]. This rule also applies to samples as a function of Ce doping, e.g., NCCO and PCCO \cite{PhysRevB.76.174512,li2019hole}, in which the $p$-MR reaches the maximum at optimal Ce doping. For LCCO, the underdoped samples without superconductivity only show $n$-MR. With increasing Ce doping, $p$-MR and superconductivity emerge simultaneously as the long-range AF order is destroyed \cite{PhysRevB.77.172503,Yu2017}. Field and doping, once again, show a resemblance in tuning the MR besides the similarity of jump in Hall number at the AF QCP. Another quantity, $B_{\max}^0$, obtained by extrapolating the $B_{\max}(T)$ to zero temperature [inset of Fig. \ref{Fig5}(d)], also shows a positive correlation with $T_{\mathrm{c}}$ [Fig. \ref{Fig5}(d)]. Larger $B_{\max}^0$ corresponds to an extended $p$-MR region, suggesting a larger field energy scale to destroy the state that causes $p$-MR. Overall, both quantities, $\delta\rho$ and $B_{\max}^0$, are intimately linked to $T_{\mathrm{c}}$. 

Notably, the $p$-MR (e.g., at 1.8 K) shows a roughly linear dependence on field. Both linear-in-temperature and linear-in-field resistivities have been discovered in LCCO from $x=0.14$ to $x=0.17$, being the signatures of a strange-metal state \cite{jin2011link,Sarkareaav6753}. We note that Sarkar {\it et al.} just reported a linear-in-temperature resistivity that extends down to $x=0.12$ (below $x_{\mathrm{AF}}$), as the resistivity upturn is suppressed at 65 T \cite{arXiv2007.12765}. For our optimally doped LCCO, the field up to 60 T cannot fully suppress the resistivity upturn [Fig. \ref{Fig2}(b)]. This means a much larger field energy scale to realize the linear-in-$T$ resistivity in LCCO with $x=0.10$, given that such feature is indeed hidden underneath the resistivity upturn. Moreover, the slope of $\rho_{xx}(B)$ for sample S1 at 1.8 K is significantly larger compared to the overdoped samples [Fig. \ref{Fig5}(b)], so a remarkably enhanced $p$-MR emerges when entering the AF region by tuning parameters $x$, $T$, or $B$. Although the micro nature of the origin of anomalous $p$-MR effect requires further elucidation, the relationships among the short-range AF order, $p$-MR and superconductivity have been vividly demonstrated.

\section{Conclusions}

In summary, we carried out a systematic magnetotransport measurement on optimally-doped cuprate LCCO films. By tuning magnetic field, the Fermi surface reconstruction is suppressed and an AF-FM phase transition happens at the critical point $B_{\mathrm{c}}$. This phase transition can be well described by holographic model. In the AF region, the MR displays a quite difference from that in long-range AF state: there is a switching from positive to negative. The $n$-MR derives from canting AF order and the $p$-MR maybe derives from topological order which is closely related to superconductivity. It is possible that similar emergent novel topological properties may exist in other electron-doped cuprates --- as the proposed mechanism for its formation is generic to the electron-doped CuO$_2$ plane. Further studies are still needed to elucidate the relationship among FS reconstruction, AF ordering, topological order and superconductivity. 

\appendix

\section{Holographic model fitting}

We consider an Einstein-Maxwell theory in $3+1$ dimensions with a negative cosmological constant and two anti-symmetry tensor fields $M_{\mu\nu}^{(1)}$ and $M_{\mu\nu}^{(2)}$. The total action reads $$S=\int d^4x\sqrt{-g}\left[R-2\Lambda-\frac14F^2-\lambda^2(L_1+L2+V_12)\right]$$with$$L_a=\frac12(dM^{(a)})^2+V(M^{(a)}),$$$a=1,2$ and $$V_{12}=\frac{k}{2}M^{(1)\mu\nu}M_{\mu\nu}^{(2)}.$$Here $L_1$ and $L_2$ are two bulk Lagrangians to describe the two different magnetic moments in staggered magnetization. The term $V_{12}$ describes the interaction between these two magnetic moments. The AF order parameter is the staggered magnetic moment, which is dual to $xy$-component $M_{xy}^{(1)}-M_{xy}^{(2)}$ in the interior. We fix the potential $V(M^{(a)})$ to be the following form $$V(M^{(a)})=M_{\mu\nu}^{(a)}M^{(a)\mu\nu}-(\epsilon_{\mu\nu\rho\sigma}M^{(a)\mu\nu}M^{(a)\rho\sigma})^2.$$Here $\epsilon_{\mu\nu\rho\sigma}$ is the Levi-Civita symbol with $\epsilon_{0123}=1$. In probe limit $\lambda\to 0$, the spacetime geometry in the interior is given by a dyonic AdS-Reissner-Nordstrom black hole which can be written as $$ds^2=-r^2f(r)dt^2+\frac{dr^2}{r^2f(r)}+r^2(dx^2+dy^2),$$$$A=\mu(1-r^{-1})dt+Bxdy,$$where $$f(r)=1-\frac{1+\mu^2+B^2}{r^3}+\frac{\mu^2+B^2}{r^4}.$$In this coordinate the boundary is at $r\to\infty$ and the horizon is at $r=1$. The chemical potential in the boundary is given by the constant $\mu$, and $B$ can be viewed as the external magnetic field of the dual boundary field theory. The temperature at boundary is given by the Hawking temperature of this black hole $$T=\frac{3-\mu^2-B^2}{4\pi}.$$The equation of motion for two anti-symmetric tensor fields reads \begin{align*}&3\nabla^\rho\nabla_{[\rho}M_{\mu\nu]}^{(a)}-4M_{\mu\nu}^{(a)}+8(\epsilon_{\gamma\tau\rho\sigma}M^{(a)\gamma\tau}M^{(a)\rho\sigma})M_{\mu\nu}^{(a)}\\&-kM_{\mu\nu}^{(b)}-F_{\mu\nu}=0.\end{align*}Here $(a,b)=(1,2)$ or $(2,1)$. Define following two quantities $\alpha$ and $\beta$ such as $$\alpha=\frac{M_{xy}^{(1)}+M_{xy}^{(2)}}{2},\quad\beta=\frac{M_{xy}^{(1)}-M_{xy}^{(2}}{2}.$$Then we have following two ordinary differential equations,
$$\alpha''+\frac{f'\alpha'}{f}-\frac{m_{\alpha\mathrm{eff}}^2}{r^2f}\alpha=\frac{B}{r^2f},$$$$\beta''+\frac{f'\beta'}{f}-\frac{m_{\beta\mathrm{eff}}^2}{r^2f}\beta=0.$$They have following asymptotic solutions as $r\to\infty$, $$\alpha\to\alpha_+r^{\frac{1+\delta_1}{2}}+\alpha_-r^{\frac{1-\delta_1}{2}}-\frac{B}{4+k},$$$$\beta\to\beta_+r^{\frac{1+\delta_2}{2}}+\beta_-r^{\frac{1-\delta_2}{2}}.$$

The detailed forms of $m_{\alpha\mathrm{eff}}^2$, $m_{\beta\mathrm{eff}}^2$, $\delta_1$, and $\delta_2$ can be found in Ref. \cite{ScheurerE3665}. The quantity $\alpha$ can be used to describe the total magnetic moment at the boundary theory and the $\beta$ can be used to describe the staggered magnetic moment of the boundary theory. The holographic description for AF state requires boundary condition $\alpha_+=\beta_+=0$. With fixed boundary chemical potential, for given external magnetic field $B$, the nonzero solution of $\beta$ can appear only when temperature is lower than a critical temperature, which corresponds to the AF transition at low temperatures.

\begin{acknowledgments}

Our work was supported by the National Key Basic Research Program of China (2017YFA0302902, 2016YFA0300301, 2017YFA0303003, and 2018YFB0704102), the National Natural Science Foundation of China (11674374, 11927808, 11834016, 118115301, 11804378 and 119611410), the Strategic Priority Research Program (B) of Chinese Academy of Sciences (XDB25000000), the Key Research Program of Frontier Sciences, CAS (QYZDB-SSW-SLH008 and QYZDY-SSW-SLH001), CAS Interdisciplinary Innovation Team, Beijing Natural Science Foundation (Z190008). This work was performed at Wuhan National High Magnetic Field Center (WHMFC), Huazhong University of Science and Technology.

\end{acknowledgments}

\bibliography{LCCO}

\end{document}